\documentclass[aps,pra,reprint,superscriptaddress,longbibliography]{revtex4-2}

\usepackage{amsmath,amssymb,amsfonts,bm,mathtools}
\usepackage{graphicx}
\usepackage{hyperref}
\usepackage{xcolor}
\usepackage{physics}

\hypersetup{
  colorlinks=true,
  linkcolor=blue,
  citecolor=blue,
  urlcolor=blue
}

\newcommand{\avg}[1]{\left\langle #1 \right\rangle}
\newcommand{\Var}{\mathrm{Var}}
\newcommand{\FI}{\mathcal{I}}
\newcommand{\CRB}{\mathrm{CRB}}

\begin{document}

\title{Differential source-basis encoding for superresolved parameter estimation in a time-reversed Young interferometer}

\author{Jianming Wen}
\email{jwen7@binghamton.edu}
\affiliation{Department of Electrical and Computer Engineering, Binghamton University, Binghamton, New York 13902, USA}

\date{\today}

\begin{abstract}
We develop a differential source-encoding protocol for local parameter estimation in a time-reversed Young interferometer, where the source plane is used not merely as a scan coordinate but as a programmable measurement basis. Two sequential positive-only source patterns implement an antisymmetric differential probe about a chosen operating point, converting the deterministicc source-coordinate response into a derivative-gradient sensing channel. In the local regime, the differential signal separates naturally into an envelope-gradient term, which is also present in noninterferometric differential sensing, and an interference-gradient term, which is specific to the time-reversed Young fringe law. This decomposition identifies the physical origin of the interferometric advantage and clarifies why it is regime dependent rather than universal. Using a shot-noise-limited Poisson model, we derive the corresponding Fisher information and Cram\'er--Rao bounds and compare the protocol with raster sampling in the same geometry and with a matched noninterferometric differential baseline. Representative numerical examples show a strong and robust gain over raster sampling, while the additional improvement from the time-reversed Young interference is parameter dependent but can be substantial in favorable regimes. The results establish the time-reversed Young geometry as a practically simple platform for programmable differential interferometric metrology.
\end{abstract}

\maketitle

\section{Introduction}
Classic Young's double-slit experiment remains one of the most familiar manifestations of optical interference, traditionally understood through first-order fringe formation on an observation screen \cite{Young1804,Goodman,MandelWolf,BornWolf}. In its standard form, a localized source illuminates the slits and the interference pattern is read out in the observation plane, where the detected fringes are shaped jointly by path interference and by the slit-diffraction envelope. That conventional picture has made Young interferometry a foundational tool in optics, but it also ties the measurement logic to observation-plane scanning or imaging \cite{Goodman,BornWolf}.

A recently proposed time-reversed Young geometry changes the logic in a fundamental way by exchanging the usual roles of source and detector \cite{Wen2025TRY,Wen2026Hybrid}. In that arrangement, a position-fixed detector replaces the conventional point source, while a laterally extended, point-addressable source plane replaces the observation screen. The resulting response is a deterministic, hybrid second-order interference law in the source coordinate, free from the usual diffraction envelope in the recorded pattern and directly programmable at the source plane \cite{Wen2025TRY,Wen2026Hybrid}. This feature makes the system more than an inverted version of ordinary Young interferometry: it defines a measurement architecture in which the source plane itself becomes the domain where the measurement basis can be engineered \cite{Wen2025TRY,Wen2026Hybrid}.

Seen from that perspective, the time-reversed Young interferometer naturally connects to computational and correlation-based imaging, including ghost imaging, differential ghost imaging, and single-pixel imaging. In those settings, structured illumination or structured detection is used to extract spatial information with one or a few detectors \cite{Shapiro2008,Bromberg2009,ErkmenShapiro2010,Ferri2010,Bina2013,Luo2012CPL,Wen2012JOSAA,Morris2015,PadgettBoyd2017,Edgar2019NP,Sun2020SPIReview,Gibson2020}. The same broad idea is relevant here, but with an important distinction: the forward response of the present system is deterministic and interferometric, and the source patterns are used not for image reconstruction from an ensemble of masks, but to synthesize a local measurement basis tailored to a parameter-estimation task. This source-plane programmability suggests a different use of the interferometer, one centered on engineered sensing channels rather than passive fringe observation. 

The problem is also closely related to modern superresolution theory \cite{Helstrom,Kay,VanTrees}. A central lesson of that field is that superresolution is most meaningfully formulated in terms of parameter-estimation performance, rather than solely in terms of point-spread-function narrowing \cite{Tsang2016PRX,NairTsang2016,NairTsang2016OE,Yang2016,Tang2016,Tsang2018,Backlund2018,Napoli2019}; see also earlier information-centered discussions of optical resolution in Refs.~\cite{Francia1955,Cox1986}. In that language, the key quantity is the Fisher information carried by the chosen measurement basis. Direct intensity measurements can become weakly informative in sub-Rayleigh regimes, whereas properly chosen alternative bases can preserve or enhance the information relevant to the unknown parameter \cite{Tsang2016PRX,NairTsang2016,NairTsang2016OE,Yang2016,Tang2016,Tsang2018,Backlund2018,Napoli2019}. Because the time-reversed Young geometry offers direct control over the source coordinate, it provides a natural platform for designing such bases directly at the measurement stage.

In this work we develop the simplest experimentally explicit realization of that idea: \emph{differential source-basis encoding} in a time-reversed Young interferometer. Rather than raster sampling one source position at a time, we use two sequential positive-only source patterns whose difference implements a local antisymmetric probe about a chosen operating point. This converts the deterministic source-coordinate response into a derivative-sensitive measurement channel for local displacement estimation. The resulting signal contains two physically distinct terms: an envelope-gradient contribution that would also appear in a generic noninterferometric differential measurement, and an interference-gradient contribution that is specific to the time-reversed Young fringe law. This decomposition identifies cleanly what is gained from differential coding itself and what is added specifically by the interferometric structure of the time-reversed geometry. 

Our aim is not to claim a universal superresolution advantage over all existing schemes. In particular, we do not argue that the present protocol generically surpass quantum-optimal mode sorting or all structured-illumination strategies. The question addressed here is more focused and, in our view, more practically relevant: when the time-reversed Young geometry is used as a programmable differential interferometer, can source-basis encoding provide a useful and distinct route to superresolved parameter estimation? We show that the anser is yes, but in a qualified and physically transparent sense. Differential encoding yields a strong improvement over naive raster sampling in the same geometry, while the traditional gain from time-reversed Young interference over a noninterferometric differential baseline depends on the operating regime and can be substantial under favorable conditions. This precisely the level of claim supported by the Fisher-information analysis and by the representative numerical comparisons developed in the manuscript.

The practical appeal of the protocol is equally important. It relies only on a fixed detector, positive-only source patterns, and differential subtraction, making it compatible with scanned spots, spatial light modulators, digital micromirror devices, or emitter arrays. In that sense, the present framework translates the unusual physics of time-reversed Young interference into an experimentally accessible architecture for programmable, fixed-detector, superresolved sensing.

This paper is organized as follows. Section \ref{sec:model} introduces the geometry and forward model. Section \ref{sec:protocol} presents the explicit differential source-encoding protocol and its positive-only implementation. Section \ref{sec:signal} derives the differential signal and isolates the interference-gradient term. Section \ref{sec:noise} develops the corresponding count statistics, Fisher information, and Cram\'er--Rao bounds. Section \ref{sec:numerics} gives representative numerical examples and quantitative comparisons with raster sampling and a noninterferometric differential baseline. Section \ref{sec:comparison} places the method in context by comparing it with standard Young interferometry, differential ghost imaging, single-pixel imaging, structured illumination, and spatial-mode demultiplexing. Section \ref{sec:conclusion} summarizes the main results and outlook. The appendixes provide additional algebra and implementation remarks.

\section{Geometry and forward model}
\label{sec:model}

\begin{figure}[t]
    \centering
    \includegraphics[width=\columnwidth]{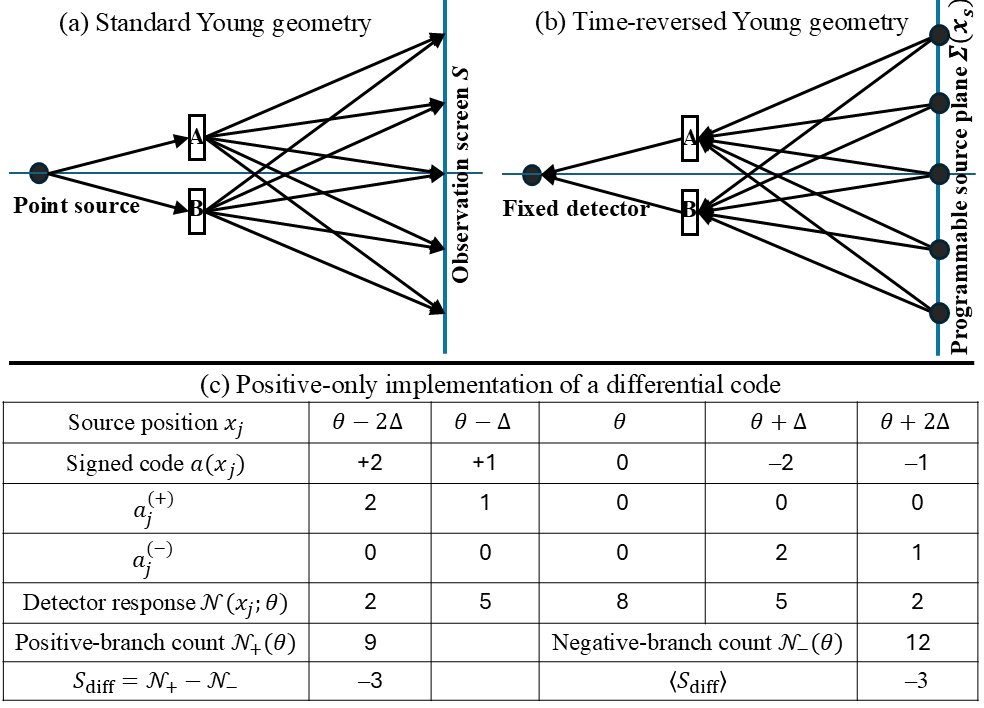}
    \caption{Schematic of the time-reversed Young interferometer and the differential source-encoding protocol. Panel (a) shows the standard Young geometry. Panel (b) shows the time-reversed geometry with a fixed detector and a programmable source plane. Panel (c) illustrates the positive-only implementation of a signed differential source code.}
    \label{fig:geometry}
\end{figure}

We consider a double-slit aperture in the plane $z=0$, with slit centers at transverse positions $x=\pm d/2$, where $d$ is the slit separation. A point detector $D$ is fixed at transverse coordinate $x_D=0$ and axial coordinate $z=-L_D$. On the opposite side of the aperture, at $z=+L_S$, lies a laterally extended and point-addressable source plane $\Sigma$  whose transverse coordinate is denoted by $x_s$. The geometry is sketched in Fig.~\ref{fig:geometry}. Unlike the conventional Young arrangement, where a fixed source illuminates the slits and the observation plane is scanned, the present time-reversed geometry keeps the detector fixed and promotes the source coordinate $x_s$ to the experimentally controlled variable \cite{Wen2025TRY,Wen2026Hybrid}.

For a source point located at $x_s$, the paraxial field arriving at slit $j\in\{A,B\}$ may be written as
\begin{equation*}
E_j(x_s)\propto \mathcal A(x_s)\exp\!\left[i\frac{k}{2L_S}(x_s-x_j)^2\right],
\end{equation*}
where $\mathcal A(x_s)$ is a source-amplitude factor, $k=2\pi/\lambda$ is the optical wavenumber, and $x_j=\pm d/2$ denotes the slit position. Propagation from slit $j$ to the detector contributes an additional phase factor
\begin{equation*}
H_j(D)\propto \exp\!\left[i\frac{k}{2L_D}(x_D-x_j)^2\right].
\end{equation*}
Because the detector is fixed at $x_D=0$, this second factor is independent of the source coordinate apart from slit-dependent constants. The source-dependent phase difference therefore comes entirely from propagation between the source plane and the two slits:
\begin{eqnarray*}
\Delta\phi(x_s)=\frac{k}{2L_S}\left[\bigg(x_s-\frac{d}{2}\bigg)^2-\bigg(x_s+\frac{d}{2}\bigg)^2\right]=-\frac{k d}{L_S}x_s.
\end{eqnarray*}
It is therefore natural to define the effective source-plane fringe spatial frequency
\begin{equation}
\kappa\equiv\frac{kd}{L_S},\label{eq:kappa}
\end{equation}
so that the interference phase varies linearly with the source coordinate. This is the basic structural feature that distinguishes the time-reversed geometry: the fringe law is written directly in the source-plane coordinate rather than in an observation-plane coordinate.

Following the time-reversed Young framework introduced in Refs.~\cite{Wen2025TRY,Wen2026Hybrid}, we model the mean detector count for a source point $x_s$ as
\begin{equation}
\mathcal N(x_s;\theta)=N_0\eta\,g(x_s-\theta)\left[1+V\cos(\kappa x_s+\phi_0)\right].
\label{eq:mu_general}
\end{equation}
Here $N_0$ is the mean photon number launched during one source-addressing interval, $\eta$ is the overall detection efficiency, $V$ is the interference visibility, $\phi_0$ is a calibrated phase offset determined by the experimental alignment. The function $g(x_s-\theta)$ is a localized response kernel centered at the unknown displacement parameter $\theta$. In physical terms, $g$ describes the spatial feature whose position is to be estimated, while the cosine factor represents the deterministic time-reversed Young modulation imposed by the interferometer.

This form is intentionally phenomenological. Its purpose is not to model every device-specific detail of a particular implementation, but to isolate the two ingredients that matter most for the sensing problem studied here: a localized source-plane response and a programmable interference law in the source coordinate. Depending on the application, the kernel $g(x_s-\theta)$ may represent a pointlike emitter broadened by finite resolution, a narrow transmissive or reflective feature in the source plane, or a localized interaction probability that is shifted by the unknown parameter $\theta$. The subsequent estimation theory depends mainly on the locality and smoothness of this kernel near the chosen operating point, not on a unique microscopic interpretation.

Equation~(\ref{eq:mu_general}) highlights two features that are central to the rest of the paper. First, the measurement is controlled directly in the source plane: varying $x_s$ changes the detected response through both the localized kernel and the source-plane interference phase. Second, the source plane is programmable, so the measurement is not restricted to sampling isolated source positions one at a time. Instead, the source coordinate can be viewed as a domain in which measurement bases are deliberately designed. This source-basis perspective motivates the differential encoding protocol developed in the next section, while its more general continuous-pattern formulation is summarized in Appendix~\ref{app:continuous}.

For illustration, Fig.~\ref{fig:response} shows the deterministic response of Eq.~(\ref{eq:mu_general}) for a Gaussian response kernel, 
\begin{equation}
    g(x_s-\theta)=\exp\left[-\frac{(x_s-\theta)^2}{2\sigma^2}\right],
\end{equation}
using representative parameter values $V=0.85$, $\sigma=1.2$, $\kappa=2.6$, and $\phi_0=\pi/5$, with $\theta=-1.2,0,1.2$. The curves show how the localized source-plane feature is translated by the displacement parameter while remaining modulated by the deterministic fringe factor. These values are chosen only to visualize the structure of the forward response; they are not intended to define a unique optimal sensing configuration.

\begin{figure}[t]
    \centering
    \includegraphics[width=\columnwidth]{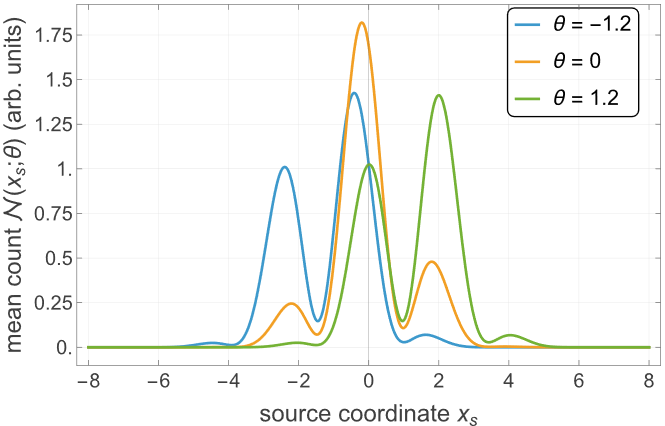}
    \caption{Deterministic time-reversed Young response as a function of the source coordinate $x_s$ for several values of the displacement parameter $\theta$. The curves illustrate Eq.~(\ref{eq:mu_general}) with a localized Gaussian response kernel multiplied by the source-coordinate fringe factor.}
    \label{fig:response}
\end{figure}

\section{Differential source-encoding protocol}
\label{sec:protocol}

The most direct use of the time-reversed geometry would be to raster sample the source plane, measuring $\mathcal N(x_s;\theta)$ one source position at a time and estimating the displacement parameter from the resulting response curve. Although straightforward, that strategy does not exploit the main experimental freedom of the platform, namely that the source plane is programmable. Our goal is instead to use that programmability to construct a local measurement basis that is directly sensitive to small changes in $\theta$. 

The basic idea is to implement a source-plane probe that is antisymmetric about a chosen operating point $x_0$. As in many differential measurement schemes, the subtraction of two closely related measurements suppresses common-mode contributions and isolates the local derivative-like part of the signal \cite{Ferri2010,Tsang2016PRX,NairTsang2016OE}. In the present setting, this converts source-plane sampling into a differential sensing channel that is naturally matched to local parameter estimation.

Because optical intensity is nonnegative, a signed source code cannot be displayed directly in a single exposure. The correct experimental realization is therefore sequential, by decomposing the signed code into positive and negative branches that are displayed one after the other; Fig.~\ref{fig:geometry}(c) gives a simple discrete example of this positive-only implementation. Let the source plane be discretized into points $x_n$, and let 
\[
\mathbf{c}=(c_1,\dots,c_M)
\]
denote a nonnegative source pattern. The corresponding mean detected count is
\begin{equation}
\mathcal N[\mathbf{c};\theta]=\sum_{n=1}^{M}c_n\,\mathcal N(x_n;\theta).\label{eq:pattern_mean}
\end{equation}
Now consider a signed code 
\[
\mathbf{a}=(a_1,\dots,a_M),\quad\sum_{n=1}^Ma_n=0.
\]
We decompose it into positive and negative branches,
\begin{equation}
a_n=a_n^{(+)}-a_n^{(-)},
\end{equation}
with 
\[
a_n^{(\pm)}\ge 0,\quad a_n^{(+)}a_n^{(-)}=0,
\]
so that the two branches have nonoverlapping support. Experimentally, one first displays the positive branch $\mathbf{a}^{(+)}$ and records the count $N_+$, then displays the negative branch $\mathbf{a}^{(-)}$ and records the count $N_-$; see Fig.~\ref{fig:geometry}(c) for an explicit discrete illustration. The differential signal is formed as
\begin{equation}
S_{\mathrm{diff}}=N_+-\gamma N_-,\label{eq:Sdiff_def}
\end{equation}
where $\gamma$ compensates any small mismatch in optical power or exposure between the two acquisitions. Its mean value is
\begin{equation}
\langle S_{\rm diff}\rangle=\mathcal N[\mathbf{a}^{(+)};\theta]-\gamma\mathcal N[\mathbf{a}^{(-)};\theta].\label{eq:Sdiff_mean}
\end{equation}
In the ideal balanced case, $\gamma=1$. When additional robustness to branch-to-branch source-power fluctuations is needed, a reference-normalized implementation may be used, as described in Appendix~\ref{app:reference}.

The simplest realization is the two-point antisymmetric probe,
\begin{equation}
x_{\pm}=x_0\pm \frac{\delta}{2},\label{eq:xpm}
\end{equation}
where $x_0$ is the operating point and $\delta$ is the source-plane probe spacing. In that case the measured differential signal reduces to
\begin{equation}
S_{\mathrm{diff}}(\theta)=N_+-N_-,\;
\avg{S_{\mathrm{diff}}}=\mathcal N(x_+;\theta)-\mathcal N(x_-;\theta).
\label{eq:Sbar_exact}
\end{equation}
This is the source analog of a finite-difference derivative measurement. It is the minimal differential probe and already captures the essential sensing mechanism studied in this paper. More general continuous or multipoint source codes (see Appendix \ref{app:reference}) are possible, but the two-point implementation is sufficient to expose the physics transparently and remains experimentally simple.

The probe spacing $\delta$ plays a central role. If $\delta$ is too small, the two measurements become nearly identical and the subtraction yields only a weak signal. If $\delta$ is too large, the probe loses locality and no longer acts as an effective differential measurement of the local response slope. The useful operating regime therefore lies between these two limits. In practice, the optimal space $\delta_{\mathrm{TRY}}^{\mathrm{opt}}$ is the value that maximizes the chosen estimation metric for the time-reversed Young protocol, such as the Fisher information or equivalently the minimum Cram\'er--Rao bound. Importantly, this spacing is not set by the wavelength alone. It is determined jointly by the width of the localized source-plane response, the a source-plane fringe period, and the selected operating point.

To maximize local sensitivity to small displacements, the operating point should be chosen near a quadrature condition \cite{Wen2025TRY} of the time-reversed Young fringe,
\begin{equation}
\kappa x_0+\phi_0=\left(m+\frac{1}{2}\right)\pi,\quad m\in\mathbb{Z},\label{eq:quadrature}
\end{equation}
for which the local slope of the cosine factor is maximal. This choice does not by itself guarantee global optimality of the full estimation protocol, but it provides the natural local working point for a derivative-sensitive measurement and is the regime of interest throughout the analysis below.

A practical measurement cycle is then straightforward. One first calibrates the deterministic source-coordinate response by scanning a suitable set of source positions and fitting Eq.~(\ref{eq:mu_general}), or an experimentally appropriate generalization of it. One next selects a working point $x_0$ near quadrature and chooses a probe spacing $\delta$. The positive branch of the differential code is displayed and the corresponding $N_+$ is recorded. The negative branch is then displayed and the count $N_-$ is recorded. Their difference gives one realization of $S_{\mathrm{diff}}$. Repeating this acquisition cycle yields a sequence of differential measurements from which the unknown displacement parameter can be inferred. In experiments where the launched source power drifts between the two branches, the same cycle can be stabilized by reference normalization, as summarized in Appendix~\ref{app:reference}. Because the detector remains fixed throughout, the method inherits the mechanical stability of fixed-detector architectures while transferring the measurement design problem to the programmable source plane. 

\begin{figure}[t]
    \centering
    \includegraphics[width=\columnwidth]{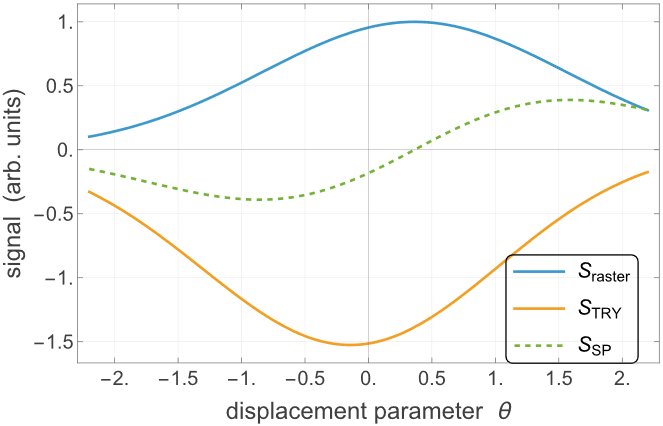}
    \caption{Comparison of a raster sample $S_{\rm raster}(\theta)$, the differential time-reversed Young signal $S_{\rm TRY}(\theta)$, and a noninterferometric differential baseline $S_{\rm SP}(\theta)$. The differential protocol converts the source-coordinate response into a derivative-sensitive channel and steepens the local dependence on the displacement parameter near the operating point.}
    \label{fig:signalcompare}
\end{figure}

Figure~\ref{fig:signalcompare} illustrates this protocol using the same Gaussian-response model and representative parameter set introduced in Sec.~\ref{sec:model}. The operating point is chosen from the quadrature condition in Eq.~(\ref{eq:quadrature}), and the probe spacing is taken as $\delta=0.8$. For the specific example in Fig.~\ref{fig:signalcompare}, we choose the nearest quadrature operating point $x_0\approx0.362$, corresponding to $m=0$, so that $x_+\approx 0.762$ and $x_1\approx -0.038$. The figure compares three signals: the raster sample \(S_{\mathrm{raster}}(\theta)=\mathcal N(x_0;\theta)\), the differential time-reversed Young signal \(S_{\mathrm{TRY}}(\theta)=\mathcal N(x_0+\delta/2;\theta)-\mathcal N(x_0-\delta/2;\theta)\), and the corresponding noninterferometric differential baseline \(S_{\mathrm{SP}}(\theta)=\mathcal N_{\mathrm{SP}}(x_0+\delta/2;\theta)-\mathcal N_{\mathrm{SP}}(x_0-\delta/2;\theta)\). Both differential signals are visibly steeper than the raster response near the operating region, confirming that differential coding provides a more sensitive local probe of small displacements. At the same time, the separation between $S_{\rm TRY}$ and $S_{\rm SP}$ shows that the time-reversed Young geometry contributes more than generic differential subtraction alone. That difference, however, should be assessed quantitatively rather than inferred from the signal shapes by eye. For that reason, the next sections develop the explicit differential signal, count statistics, Fisher information and Cram\'{e}r--Rao bounds.

\section{Differential signal and its physical content}
\label{sec:signal}

Substituting the forward model of Eq.~(\ref{eq:mu_general}) into the two-point differential measurement of Eq.~(\ref{eq:Sbar_exact}), we obtain the mean differential signal
\begin{align}
\bar S(\theta)
&\equiv \avg{S_{\mathrm{diff}}} \nonumber\\
&=N_0\eta\Bigg\{g\left(x_0+\frac{\delta}{2}-\theta\right)
\left[1+V\cos\left(\kappa x_0+\frac{\kappa\delta}{2}+\phi_0\right)\right] \nonumber\\
&\quad -g\left(x_0-\frac{\delta}{2}-\theta\right)
\left[1+V\cos\left(\kappa x_0-\frac{\kappa\delta}{2}+\phi_0\right)\right]\Bigg\}.
\label{eq:Sbar_full}
\end{align}
This expression is exact for the two-point probe. It shows that the differential output is shaped jointly by the local variation of the source-feature envelope $g$ and by the phase-sensitive fringe response of the time-reversed Young geometry. In that sense, the protocol is not merely subtractive; it is subtractive and interferometric at the same time.

To expose the local sensing structure, we consider the regime in which the probe spacing $\delta$ is small compared with the scale over which the response varies appreciably. The two-point difference then becomes a finite-difference approximation to the source-coordinate derivative,
\begin{equation}
\bar S(\theta)\approx\delta\,\partial_x\mathcal N(x_0;\theta). \label{eq:Sbar_derivative}
\end{equation}
Using Eq.~(\ref{eq:mu_general}), one finds
\begin{align}
\partial_x\mathcal N(x;\theta)&=N_0\eta\Big\{g'(x-\theta)\left[1+V\cos(\kappa x+\phi_0)\right] \nonumber\\
&\qquad\quad-\kappa V\,g(x-\theta)\sin(\kappa x+\phi_0)\Big\}.\label{eq:dmu_dx}
\end{align}
Evaluating this at the operating point $x_0$ gives
\begin{align}
\bar S(\theta)&\approx N_0\eta\,\delta\Big\{
g'(x_0-\theta)\left[1+V\cos(\kappa x_0+\phi_0)\right] \nonumber\\
&\qquad\qquad
-\kappa V\,g(x_0-\theta)\sin(\kappa x_0+\phi_0)
\Big\}.\label{eq:Sbar_firstorder}
\end{align}
Equation~(\ref{eq:Sbar_firstorder}) already shows the central mechanism of the protocol: differential source encoding converts the forward response into a derivative-sensitive measurement channel. The first factor comes from the local slope of the source-feature envelope, whereas the second arises from the slope of the source-plane interference fringe itself. These two contributions play qualitatively different roles, and their relative importance depends on the chosen operating point.

The most informative local operating point is a quadrature point of Eq.~(\ref{eq:quadrature}), for which
\begin{equation*}
\cos(\kappa x+\phi_0)=0,\quad \sin(\kappa x_0+\phi_0)=\pm 1.
\end{equation*}
At quadrature, Eq.~(\ref{eq:Sbar_firstorder}) simplifies to
\begin{equation}
\bar S(\theta)\approx N_0\eta\,\delta\left[g'(x_0-\theta)\mp\kappa V\,g(x_0-\theta)\right].\label{eq:Sbar_quadrature}
\end{equation}
Equation~(\ref{eq:Sbar_quadrature}) is the central analytical result of the present work. It shows that the differential signal contains two physically distinct terms. The first, $g'(x_0-\theta)$, is an \emph{envelope-gradient term}. It would also appear in a generic noninterferometric differential (single-pixel) measurement, since it reflects only the local slope of the source-feature response. The second, $\mp\kappa V\,g(x_0-\theta)$, is an \emph{interference-gradient term}. It is specific to the time-reversed Young fringe law and vanishes if the cosine modulation in Eq.~(\ref{eq:mu_general}) is removed. This is separation makes it possible to identify precisely what the time-reversed geometry contributes beyond differential sampling alone. 

This decomposition also clarifies the design scales of the problem. The source-feature width is set by the kernel $g$, often characterized by a scale $\sigma$. The source-plane fringe period is
\begin{equation}
\Lambda_s=\frac{2\pi}{\kappa}=\frac{\lambda L_S}{d}.\label{eq:LambdaS}
\end{equation}
The probe spacing $\delta$ should therefore be judged relative to both $\sigma$ and $\Lambda_s$. In particular, the optimal differential spacing is not meaningfully described as a universal multiple of the optical wavelength. A more useful characterization is through the dimensionless ratios
\[
\frac{\delta_{\mathrm{TRY}}^{\mathrm{opt}}}{\sigma},\quad\frac{\delta_{\mathrm{TRY}}^{\mathrm{opt}}}{\Lambda_s},
\]
which reflect, respectively, the locality of the source probe relative to the feature width and the phase sensitivity of the probe relative to the source-plane fringe scale. This viewpoint will be used throughout the numerical analysis below.

For the Gaussian kernel used in the representative examples, the corresponding explicit formulas for the signal, its parameter derivative, and the exact branch-count Fisher information are given in Appendix~\ref{app:gaussian}. Those formulas are especially useful for numerical optimization because they show directly how the envelope-gradient and interference-gradient contributions combine in a concrete and experimental relevant model.

\section{Count statistics, Fisher information, and Cram\'er--Rao bounds}
\label{sec:noise}

We now quantify the displacement-estimation performance of the differential source-encoding protocol. Because the measurement is implemented through two sequential positive-only branches, the fundamental observables are the two branch counts themselves, not only their difference. We assume shot-noise-limited operation, so that the two branch counts are statistically independent Poisson random variables with means
\begin{equation}
N_{\pm}\sim\mathrm{Pois}(\mathcal N_{\pm}),\quad
\mathcal N_{\pm}\equiv \mathcal N\left(x_0\pm\frac{\delta}{2};\theta\right),
\end{equation}
where $\mathcal N(x_s;\theta)$ is the forward response of Eq.~(\ref{eq:mu_general}).This model is the natural one for photon-counting operation with sequentially displaced source patterns.

For the simple differential estimator $S_{\rm diff}=N_+-N_-$, the mean is the differential signal (9) derived in Sec.~\ref{sec:protocol}, and, by independence of the two Poisson branches, the variance is
\begin{equation}
\Var(N_{\pm})=\mathcal N_{\pm},\quad
\Var(S_{\mathrm{diff}})=\mathcal N_+ + \mathcal N_-. \label{eq:var_diff}
\end{equation}
This immediately gives a useful local slope-to-noise sensitivity metric,
\begin{equation}
\mathcal{M}(\theta)=\frac{|\partial_\theta \bar S(\theta)|}{\sqrt{\Var(S_{\mathrm{diff}})}},
\label{eq:sens_metric}
\end{equation}
which measures how rapidly the mean differential signal changes relative to the shot-noise level. Although $\mathcal M(\theta)$ is convenient for visualizing local sensitivity, it is not by itself the fundamental bound on estimation precision. For that purpose, the Fisher information of the full counting statistics is the correct quantity.

Because the two branch counts are independent Poisson variables, the exact Fisher information carried by the differential time-reversed Young protocol is
\begin{equation}
\FI_{\mathrm{TRY}}(\theta)=\frac{\left(\partial_\theta\mathcal N_+\right)^2}{\mathcal N_+}+\frac{\left(\partial_\theta\mathcal N_-\right)^2}{\mathcal N_-}.\label{eq:FI_exact}
\end{equation}
The corresponding Cram\'er--Rao bound for any unbiased estimator of $\theta$ is
\begin{equation}
\Delta\theta_{\CRB}\ge \frac{1}{\sqrt{\FI_{\mathrm{TRY}}(\theta)}}.
\label{eq:CRB_exact}
\end{equation}
Equations~(\ref{eq:FI_exact}) and (\ref{eq:CRB_exact}) provide the primary quantitative metrics used throughout the rest of the paper. In particular, they allow the present protocol to be compared on equal footing with raster sampling and with noninterferometric differential baselines.

The local signal slope entering Eq.~(\ref{eq:sens_metric}) follows directly from Sec.~\ref{sec:signal}. Differentiating Eq.~(\ref{eq:Sbar_firstorder}) with respect to $\theta$ gives
\begin{align}
\partial_\theta \bar S(\theta)&\approx-N_0\eta\,\delta\Big\{
g''(x_0-\theta)\left[1+V\cos(\kappa x_0+\phi_0)\right]\nonumber\\
&\qquad\quad -\kappa V\,g'(x_0-\theta)\sin(\kappa x_0+\phi_0)\Big\}.\label{eq:dS_dtheta_general}
\end{align}
At a quadrature operating point [Eq.~(\ref{eq:quadrature})], this simplifies to
\begin{equation}
\partial_\theta \bar S(\theta)\approx-N_0\eta\,\delta\left[g''(x_0-\theta)\mp \kappa V\,g'(x_0-\theta)\right].\label{eq:dS_dtheta_quadrature}
\end{equation}
Together, Eqs.~(\ref{eq:Sbar_quadrature}) and (\ref{eq:dS_dtheta_quadrature}) show that the protocol is intrinsically local: it probes not only the amplitude of the source-feature response near $x_0$, but also its first- and second-derivative structure, while simultaneously exploiting the local slope of the source-plane interference fringe. This is the estimation-theoretic counterpart of the signal decomposition developed in Sec.~\ref{sec:protocol}.

For comparison, consider first a raster measurement in the same time-reversed Young geometry. In that case one samples a single source position $x_0$ and uses the corresponding count 
\[
\mathcal N_{\rm raster}(\theta)\equiv\mathcal N(x_0;\theta).
\]
Its Fisher information is
\begin{equation}
\FI_{\mathrm{raster}}(\theta)=\frac{\left[\partial_\theta\mathcal N(x_0;\theta)\right]^2}{\mathcal N(x_0;\theta)}.\label{eq:FI_raster}
\end{equation}
This is the appropriate same-platform baseline because it retains the same time-reversed interferometric response while removing the differential source-basis construction. The comparison between $\mathcal I_{\rm TRY}$ and $\mathcal I_{\rm raster}$ hence isolates the gain obtained from differential encoding itself.

A second baseline is obtained by removing the time-reversed Young fringe modulation while preserving the same two-branch differential acquisition logic. Specifically, we define the noninterferometric single-pixel model
\begin{equation}
\mathcal N_{\mathrm{SP}}(x_s;\theta)=N_0\eta\,g(x_s-\theta),\label{eq:mu_sp}
\end{equation}
which differs from Eq.~(\ref{eq:mu_general}) only by the absence of the cosine term. The two branch means are then
\[
\mathcal N_{\rm SP,\pm}\equiv\mathcal N\left(x_0\pm\frac{\delta}{2};\theta\right),
\]
and the corresponding Fisher information is
\begin{equation}
\FI_{\mathrm{SP}}(\theta)=\frac{\left(\partial_\theta\mathcal N_{\mathrm{SP},+}\right)^2}{\mathcal N_{\mathrm{SP},+}}+\frac{\left(\partial_\theta\mathcal N_{\mathrm{SP},-}\right)^2}{\mathcal N_{\mathrm{SP},-}}.\label{eq:FI_sp}
\end{equation}
This baseline is especially important because it isolates what is gained specifically from the time-reversed Young interference, rather than from differential sampling alone. In other words, the comparison $\FI_{\rm TRY}$ versus $\mathcal I_{\rm SP}$ directly tests the added value of the interference-gradient contribution identified in Sec.~\ref{sec:signal}.

For general kernels $g$, the Fisher information may be evaluated numerically from Eqs.~(\ref{eq:FI_exact}), (\ref{eq:FI_raster}), and (\ref{eq:FI_sp}). For the Gaussian model used in the representative examples below, explicit closed-form expressions for the branch means and the exact Fisher information are collected in Appendix~\ref{app:gaussian}. Those formulas are used in Sec.~\ref{sec:numerics} to map the dependence of the estimation performance on the probe spacing $\delta$, the fringe scale $\kappa$, and the chosen operating point.

\section{Representative numerical examples}
\label{sec:numerics}

We now turn to representative numerical examples based on the Gaussian model summarized in Appendix~\ref{app:gaussian}. Unless otherwise stated, we use
\begin{equation*}
\sigma=1.2,\quad V=0.85,\quad \kappa=2.6,\quad \phi_0=\frac{\pi}{5}, \label{eq:numparams}
\end{equation*}
and compare the differential time-reversed Young protocol with the two baselines introduced in Sec.~\ref{sec:protocol}: raster sampling in the same geometry and the noninterferometric differential protocol. Again, these values are chosen only to illustrate the structure of the problem and do not represent a unique optimized experimental design. 

\begin{figure}[t]
    \centering
    \includegraphics[width=\columnwidth]{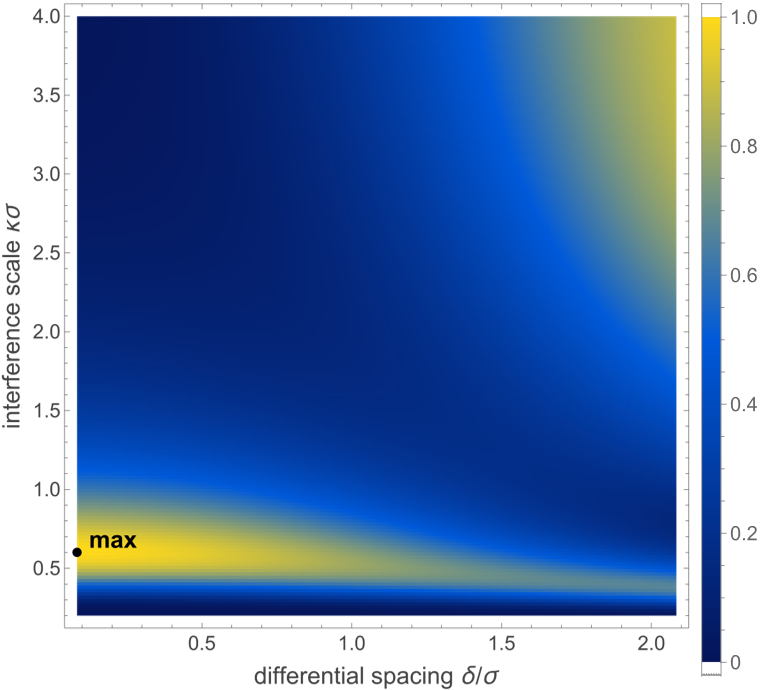}
    \caption{Local Fisher information of the differential time-reversed Young protocol as a function of the normalized differential spacing $\delta/\sigma$ and the dimensionless interference parameter $\kappa \sigma$. The marked point indicates the maximum within the plotted parameter range.}
    \label{fig:fishermap}
\end{figure}

Figure~\ref{fig:fishermap} provides a global view of the performance of the differential time-reversed Young protocol by mapping the local Fisher information $\FI_{\mathrm{TRY}}$ as a function of the normalized probe spacing $\delta/\sigma$ and the dimensionless interference parameter $\kappa\sigma$. For each point in the map, the operating point $x_0$ is chosen from the quadrature condition in Eq.~(\ref{eq:quadrature}), and the Fisher information is evaluated at a fixed small displacement offset in order to characterize local sensitivity. The figure therefore serves as a design map: it shows how the estimation performance depends jointly on the locality of the differential probe and on the source-plane fringe scale relative to the feature width.

Several general trends are evident. When $\delta/\sigma$ is too small, the two probe branches sample nearly the same part of the response, so the subtraction carries little information. When $\delta/\sigma$ is too large, the probe loses locality and no longer functions as an efficient differential measurement of the local response slope. Between these two limits lies an optimal region. The map also shows that the performance depends nontrivially on $\kappa\sigma$: the interferometric contribution is most useful when the source-plane fringe scale is well matched to the localized response. In this sense, Fig.~\ref{fig:fishermap} is better viewed as a joint design landscape than as a one-parameter optimization curve.

To make the comparison more explicit, we next evaluate two representative quadrature operating points. We first take the nominal displacement value $\theta_0=0$ and choose the nearest quadrature point,
\begin{equation*}
x_0=\frac{\pi/2-\phi_0}{\kappa}\approx 0.3625.
\end{equation*}
Optimizing over the range $0.1\le \delta \le 2.5$ gives
\[
\delta_{\mathrm{TRY}}^{\mathrm{opt}}\approx 2.50,
\]
with nearly the same optimum for the noninterferometric differential baseline. At this operating point, the differential time-reversed Young protocol yields
\[
\mathcal{M}_{\mathrm{TRY}}\approx 0.866,\quad\FI_{\mathrm{TRY}}\approx 0.818,\quad \Delta\theta_{\mathrm{TRY}}\approx 1.106.
\]
By contrast, raster sampling in the same geometry gives
\[
\mathcal{M}_{\mathrm{raster}}\approx 0.246,\quad\FI_{\mathrm{raster}}\approx 0.061,\quad \Delta\theta_{\mathrm{raster}}\approx 4.064.
\]
Thus, relative to raster sampling, the differential protocol improves the local sensitivity by about a factor of 3.5, increases the Fisher information by about a factor of 13.5, and reduces the Cram\'er--Rao bound by about a factor of 3.7. The matched noninterferometric differential baseline yields
\[
\mathcal{M}_{\mathrm{SP}}\approx 0.855,\quad\FI_{\mathrm{SP}}\approx 0.797,\quad \Delta\theta_{\mathrm{SP}}\approx 1.120,
\]
showing that at this particular operating point the dominant gain comes from differential coding itself, while the additional contribution from the time-reversed Young interference is present but comparatively mild.

A second example is more favorable to the interferometric geometry. Choosing the next quadrature point,
\begin{equation*}
x_0=\frac{\pi/2-\phi_0+2\pi}{\kappa}\approx 2.779,
\end{equation*}
one finds \[
\delta_{\mathrm{TRY}}^{\mathrm{opt}}\approx 1.47,\quad \delta_{\mathrm{SP}}^{\mathrm{opt}}\approx 2.42.
\]
At this operating point, the time-reversed protocol gives 
\[
\mathcal{M}_{\mathrm{TRY}}\approx 0.909,\quad \FI_{\mathrm{TRY}}\approx 0.867,\quad \Delta\theta_{\mathrm{TRY}}\approx 1.074,
\]
whereas the noninterferometric differential baseline gives 
\[
\mathcal{M}_{\mathrm{SP}}\approx 0.690, \quad\FI_{\mathrm{SP}}\approx 0.536,\quad \Delta\theta_{\mathrm{SP}}\approx 1.366.
\]
The interferometric contribution therefore improves the local sensitivity by about $32\%$, increases the Fisher information by about $62\%$, and reduces the Cram\'er--Rao bound by about $27\%$ relative to the noninterferometric differential baseline. At the same operating point, raster sampling gives 
\[
\mathcal{M}_{\mathrm{raster}}\approx 0.505,\quad \FI_{\mathrm{raster}}\approx 0.255, \quad\Delta\theta_{\mathrm{raster}}\approx 1.981,
\]
so the differential time-reversed scheme still outperforms raster sampling by a substantial margin.

These two examples sharpen the main numerical message of this work. The most robust improvement comes from replacing raster sampling by a local antisymmetric differential probe. The time-reversed Young interference then adds a second contribution that is not universal but can be significant in favorable regimes. This is precisely the behavior suggested by Eq.~(\ref{eq:Sbar_quadrature}): the interference-gradient term enhances the envelope-gradient term, but does not replace it. The time-reversed geometry should therefore be understood not as a universally dominant mechanism on its own, but as an interferometric enhancement of an already effective differential sensing architecture. 

It is also useful to interpret the optimized probe spacing in the dimensionless form. In the second example,
\begin{equation*}
\frac{\delta_{\mathrm{TRY}}^{\mathrm{opt}}}{\sigma}\approx \frac{1.47}{1.2}\approx 1.23,
\end{equation*}
so the optimal probe spacing is of the same order as the width of the localized source feature. At the same time,
\begin{equation*}
\frac{\delta_{\mathrm{TRY}}^{\mathrm{opt}}}{\Lambda_s}=
\frac{\delta_{\mathrm{TRY}}^{\mathrm{opt}}\kappa}{2\pi}\approx 0.61,
\end{equation*}
showing that it is also a substantial fraction of the source-plane fringe period. This interpretation is more informative than quoting $\delta_{\mathrm{TRY}}^{\mathrm{opt}}$ as an isolated dimensional number, because the relevant design scales are set jointly by the source-feature width and by the fringe period in the programmable source plane.

\begin{figure*}[t]
    \centering
    \includegraphics[width=\textwidth]{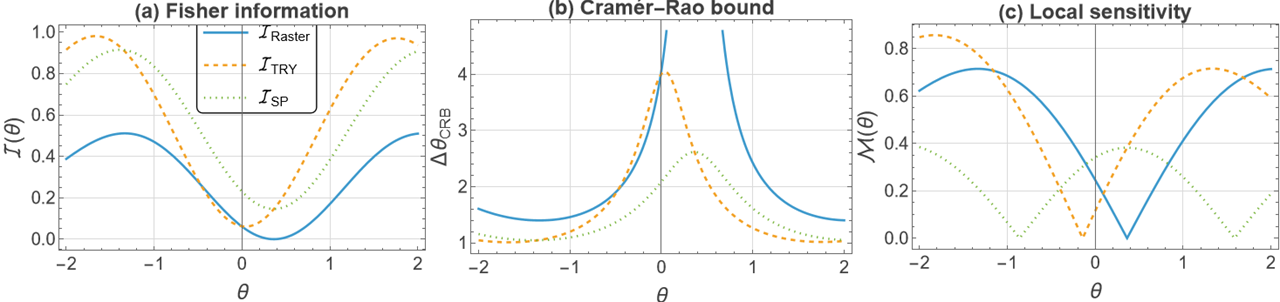}
    \caption{Comparison of the differential time-reversed Young protocol with raster sampling and the noninterferometric differential baseline for the same Gaussian-response model and representative parameter set used in the text. Panel (a) shows the Fisher information, panel (b) the corresponding Cram\'er--Rao bound, and panel (c) the local sensitivity metric $|\partial_\theta S|/\sqrt{\mathrm{Var}(S)}$, all plotted as functions of the displacement parameter $\theta$. The time-reversed Young differential protocol and the noninterferometric differential baseline both outperform raster sampling over the main operating region, while the relative advantage of the time-reversed Young protocol over the noninterferometric differential case is parameter dependent and must therefore be assessed quantitatively rather than inferred from the signal shapes alone.}
    \label{fig:comparisonpanel}
\end{figure*}

Figure~\ref{fig:comparisonpanel} presents the same comparison in a complementary form by plotting the Fisher information, the corresponding Cram\'er--Rao bound, and the local sensitivity metric as functions of the displacement parameter $\theta$. Both differential protocols outperform raster sampling over the main operating region, while the relative advantage of the time-reversed Young protocol over the noninterferometric differential baseline remains parameter dependent. This comparison is important because it reinforces a central point of the paper: visual steepness of the signal curves alone is not sufficient to establish metrological advantage. The relevant comparison must be made through the full counting statistics and the resulting estimation metrics. 

Taken together, Figs.~\ref{fig:fishermap} and \ref{fig:comparisonpanel} show that the present protocol is neither a simple reformulation of raster scanning nor a universal superior interferometric strategy. Its practical strength comes from the combination of three elements: fixed-detector operation, programmable differential source coding, and the additional interference-gradient contribution supplied by the time-reversed Young geometry. When these ingredients are jointly matched to the operating regime, the method provides a simple and quantitatively effective route to local superresolved parameter estimation.

\section{Comparison with related schemes}
\label{sec:comparison}

The present protocol can be further best understood by comparison with several neighboring classes of optical measurement schemes. The relevant distinctions are not only geometric but also operational: what quantity is controlled, what quantity is measured, and how the measurement basis is constructed. From that perspective, the time-reversed Young protocol occupies a specific niche: it is a fixed-detector, source-programmable, differential interferometric method for local parameter estimation.

The first comparison is with standard Young interferometry. In the conventional geometry [Fig.~\ref{fig:geometry}(a)], the source is fixed, the observation plane is scanned or imaged, and the recorded pattern is shaped by both path interference and the slit-diffraction envelope \cite{Young1804,Goodman,BornWolf,MandelWolf}. In the time-reversed geometry, by contrast, the detector is fixed and the source plane becomes the programmable degree of freedom \cite{Wen2025TRY,Wen2026Hybrid}. The distinction is thus not merely that source and detector are interchanged. More importantly, the source plane is used here to define a \emph{local differential measurement basis}, rather than simply to generate a fringe pattern. In that sense, the central departure from ordinary Young interferometry is the shift from passive fringe observation to active source-basis design.

The second comparison is with differential ghost imaging. Differential ghost imaging improves the signal-to-noise ratio of correlation-based bucket-type measurements by subtracting suitably normalized signals so as to suppress common-mode backgrounds \cite{Ferri2010,Bina2013}. That subtraction logic is closely related to the present protocol, and the positive-only two-branch implementation used here is conceptually similar to that normalization strategy. The forward model, however, is fundamentally different. In differential ghost imaging, the image is reconstructed from intensity correlations accumulated over an ensemble of illumination patterns. In present scheme, the forward law is deterministic and interferometric, and the source patterns are chosen deliberately to approximate a local antisymmetric derivative probe around a calibrated quadrature point. The method is therefore closer in spirit to a \emph{differential interferometric single-pixel measurement} than to conventional ghost imaging.

The third comparison is with single-pixel imaging and computational ghost imaging more broadly. In those settings, structured illumination or structured detection is combined with one or a few detectors to infer spatial information \cite{Shapiro2008,Bromberg2009,ErkmenShapiro2010,Luo2012CPL,Wen2012JOSAA,Morris2015,PadgettBoyd2017,Edgar2019NP,Sun2020SPIReview,Gibson2020}. The present geometry clearly belongs to that large family because the source-plane pattern defines the measurement basis. What distinguishes it is that the basis is not introduced only for compression, reconstruction, or correlation averaging. Instead, it is chosen to realize a local derivative-sensitive sensing channel around a calibrated operating point. In the absence of the time-reversed Young fringe modulation, the same two-branch differential protocol reduces to a noninterferometric single-pixel baseline. The time-reversed geometry then adds the interference-gradient contribution identified in Sec.~\ref{sec:signal}, which provides an additional metrological resource beyond differential subtraction alone. The method is therefore better viewed as a source-programmable differential single-pixel interferometer than as generic single-pixel imaging.

This distinction is also borne out quantitatively. In the representative example of Sec.~\ref{sec:numerics} with the second quadrature operating point, the optimized time-reversed Young protocol yields $\FI_{\rm TRY}\approx0.867$ and $\Delta\theta_{\rm TRY}\approx1.074$, whereas the matched noninterferometric differential baseline gives $\FI_{\rm SP}\approx0.536$ and $\Delta\theta_{\rm SP}\approx1.366$. Thus,
\[
\frac{\FI_{\rm TRY}}{\FI_{\rm SP}}\approx1.62,\quad\frac{\Delta\theta_{\rm SP}}{\Delta\theta_{\rm TRY}}\approx1.27.
\]
At the first representative operating point, by contrast, the corresponding gains are much smaller. This again shows that the time-reversed Young interference is not the sole origin of the protocol’s usefulness, but rather an additional contribution whose benefit depends on the operating regime.

The fourth comparison is with structured illumination microscopy. Linear structured illumination enhances recoverable spatial bandwidth by mixing otherwise inaccessible spatial frequencies into the passband, and under suitable conditions can improve lateral resolution by roughly a factor of two \cite{Gustafsson2000,Gustafsson2009,Saxena2015,Schermelleh2019}. The present protocol does not pursue that objective. It is not a full-field bandwidth-extension method, nor is it designed to reconstruct an image over an extended spatial domain. Rather, it is a local parameter-estimation scheme in which the relevant performance metric is the Fisher information for a displacement parameter. Its superresolution content is hence estimation-theoretic rather than image-bandwidth based. For that reason, comparisons with structured illumination or STED \cite{Saxena2015,Schermelleh2019,Hell1994} should be understood at the level of general superresolution philosophy, not as direct like-for-like benchmarking. 

The strongest benchmark is spatial-mode demultiplexing and related mode-sorting strategies \cite{Tsang2016PRX,NairTsang2016,NairTsang2016OE,Yang2016,Tang2016,Tsang2018,Backlund2018,Napoli2019,Zhou2019,Wadood2021}. Those methods were developed precisely to preserve Fisher information in sub-Rayleigh parameter-estimation problems and, in idealized settings, can attain the relevant quantum limits. The present work does not compete with that class of schemes on fundamental optimality. Indeed, the current protocol uses only two sequential source-plane branches and a fixed detector, so it should be viewed as a deliberately simple and experimentally accessible measurement architecture rather than as a quantum-optimal one. The appropriate question is therefore not whether it universally surpasses mode sorting, but whether it offers a useful intermediate strategy in settings where fully optimized mode demultiplexing is unavailable, unnecessary, or experimentally cumbersome. In that practical sense, the present method occupies a different point in the trade-off between implementation simplicity and estimation performance.

This practical positioning can be summarized through the hierarchy of comparisons developed in Secs.~\ref{sec:protocol} and \ref{sec:comparison}. Relative to raster sampling in the same time-reversed geometry, the gain from differential source coding can be large. For example, at the first representative operating point,
\[
\frac{\FI_{\rm TRY}}{\FI_{\rm raster}}\approx\frac{0.818}{0.061}\approx13.5,
\]
while at the second operating point,
\[
\frac{\FI_{\rm TRY}}{\FI_{\rm raster}}\approx\frac{0.867}{0.255}\approx 3.4,
\]
Thus the clearest and most robust advantage of the present framework is not over every conceivable superresolution architecture, but over naive point-by-point sampling in the same physical platform. The comparison with the noninterferometric differential baseline then isolates the more specific role of the time-reversed Young interference itself.

Taken together, these comparisons clarify the proper scope of the present claim. The protocol is not conventional Young interferometry, because the source plane is used as a programmable measurement basis. It is not conventional ghost imaging, because the forward response is deterministic rather than correlation reconstructed. It is not merely generic single-pixel imaging, because the explicit interference-gradient term gives a distinct interferometric contribution beyond differential subtraction alone. It is not a bandwidth-extension method like structured illumination microscopy, because its superresolution content is local and estimation theoretic. And it is not intended to replace quantum-optimal mode sorting, because its emphasis is experimental simplicity rather than absolute optimality. Its real niche is narrow but well defined: a practically accessible, source-programmable differential interferometric strategy that can deliver substantial gains over raster sampling and, in favorable regimes, meaningful additional gains over noninterferometric differential sensing.

\section{Conclusion}
\label{sec:conclusion}

We have developed a theory of differential source encoding in a time-reversed Young interferometer and shown that the source place can be used not merely as a scanning coordinate, but as a programmable measurement basis for local parameter estimation. In this architecture, antisymmetric source coding converts the deterministic source-coordinate response into a derivative-sensitive sensing channel, thereby turning the time-reversed Young geometry into a programmable differential interferometer. 

The central analytical result is that, near a quadrature operating point, the differential signal separates naturally into two physically distinct contributions: an envelope-gradient term and an interference-gradient term. The first reflects the local slope of the underlying source-feature response and is already present in a noninterferometric differential protocol. The second is specific to the time-reversed Young fringe law and identifies precisely what the interferometric geometry adds beyond differential subtraction alone. This decomposition provides a compact physical interpretation of the protocol and clarifies why its metrological advantage is intrinsically regime dependent rather than universal.

The numerical examples reinforce this picture. The clearest and most robust gain comes from replacing raster sampling by a local antisymmetric differential probe in the same time-reversed platform. In the representative examples studied here, the differential protocol improves the Fisher information over raster sampling by factors of about 13.5 and 3.4 at two different quadrature operating points, with corresponding reductions of the Cram\'er--Rao bound by factors of about 
3.7 and 1.8, respectively. Relative to the matched noninterferometric differential baseline, the added interferometric contribution is modest in some regimes but can be substantial in others; in the more favorable example, it increases the Fisher information by about 62$\%$ and reduces the Cram\'er--Rao bound by about 27$\%$. These results make clear that the strongest improvement arises from differential source-basis design itself, while the time-reversed Young interference can provide a meaningful additional enhancement when the operating point and fringe scale are well matched.

The practical appeal of the protocol lies in its simplicity and flexibility. It requires only a fixed detector, positive-only source patterns, and differential subtraction, and is therefore compatible with scanned spots, digital micromirror devices, spatial light modulators, or emitter arrays. Because the detector remains fixed, the method naturally inherits the stability of fixed-detector architectures, while the measurement design is transferred to the programmable source plane. In this sense, the time-reversed Young geometry offers a distinctive experimental route to differential sensing that combines source-basis control with interferometric enhancement.

The superresolution content of the present work should be interpreted in the parameter-estimation sense. The protocol is not intended to replace quantum-optimal mode sorting or full-field bandwidth-extension methods such as structured illumination. Its contribution is different and more targeted: it shows that the time-reversed Young geometry can be used as a practically accessible platform for local superresolved metrology, in which the measurement basis is engineered directly at the source plane and implemented through a simple two-branch differential acquisition.

More broadly, the present framework suggests that time-reversed interferometric geometries may be useful not only for reproducing familiar fringe laws in unconventional coordinates, but also for designing new measurement architectures in which programmability and interferometric structure are exploited together. Extensions to continuous source-pattern families, adaptive source-basis design, multipoint differential codes, and task-specific computational sensing strategies appear particularly promising. From that perspective, the main significance of this work is not only the specific protocol analyzed here, but the broader recognition that the source plane in a time-reversed interferometer can serve as a flexible and quantitatively useful measurement-design space.

\section*{Acknowledgments}

The authors are grateful to Drs.~Bing He, Yanhua Zhai, and Da Zhang, and Ms.~Yu Zhang for useful discussions related to the results reported in this work. This work was partially supported by startup funds from Binghamton University and NSF ExpandQISE 2329027.

\appendix

\section{Continuous-pattern formulation}
\label{app:continuous}

The two-point code discussed in the main text is the simplest differential probe, but the same framework extends immediately to continuous source patterns. Let $a(x)$ be a signed source code defined on the source plane and satisfying
the zero-mean condition
\begin{equation}
\int a(x)\,\dd x = 0.
\end{equation}
The corresponding mean response is
\begin{equation}
\bar S_a(\theta)=\int a(x)\,\mathcal N(x;\theta)\,\dd x,\label{eq:Sa}
\end{equation}
where $\mathcal N(x;\theta)$ is the source-coordinate response introduced in Eq.~(\ref{eq:mu_general}).

When $a(x)$ is antisymmetric about an operating point $x_0$ and sufficiently localized, Eq.~(\ref{eq:Sa}) reduces to a derivative-weighted measurement of the local response near $x_0$, namely,
\[
\bar S_a(\theta)\propto\partial_x \mathcal N(x_0;\theta),
\]
up to a code-dependent prefactor. In this sense, the two-point differential probe used in the main text is simply the minimal discrete realization of a broader source-basis strategy in which the source pattern is designed to extract local derivative information.

This continuous viewpoint is useful for two reasons. First, it places the present protocol in a more general source-basis framework that includes multipoint and smoothly varying differential codes, rather than only the two-point antisymmetric pair. Second, it makes contact with orthogonal pattern families used in single-pixel and computational imaging, while emphasizing that the objective here is not full image reconstruction but local parameter estimation. It therefore suggests a natural route to adaptive source-basis design, in which the source pattern is optimized for a specific estimation task or operating regime.

\section{Gaussian kernel and explicit formulas}
\label{app:gaussian}

For the Gaussian response kernel
\begin{equation}
g(u)=\exp\!\left(-\frac{u^2}{2\sigma^2}\right),
\end{equation}
the first two derivatives are
\begin{equation}
g'(u)=-\frac{u}{\sigma^2}g(u),\quad
g''(u)=\left(\frac{u^2}{\sigma^4}-\frac{1}{\sigma^2}\right)g(u).
\end{equation}
Let
\[
u\equiv x_0-\theta.
\]
Substituting Eq.~(B2) into teh quadrature expression of Eq.~(\ref{eq:Sbar_quadrature}), the approximate mean differential signal becomes
\begin{equation}
\bar S(u)\approx N_0\eta\,\delta\,g(u)\left[-\frac{u}{\sigma^2}\mp \kappa V\right].\label{eq:Sgauss}
\end{equation}
Likewise, differentiating with respect to $\theta$ gives
\begin{equation}
\partial_\theta \bar S(u)\approx N_0\eta\,\delta\,g(u)
\left[\frac{1}{\sigma^2}-\frac{u^2}{\sigma^4}\mp \kappa V\frac{u}{\sigma^2}
\right]. \label{eq:dSgauss}
\end{equation}
Equations~(\ref{eq:Sgauss}) and (\ref{eq:dSgauss}) make the two-channel structure especially transparent. The terms involving only powers of $u/\sigma$ arise from the envelope-gradient contribution, while the terms proportional to $\kappa V$ arise from the time-reversed Young interference. The Gaussian model therefore provides a convenient analytic setting in which the purely differential and specifically interferometric parts of the response can be examined separately and then recombined quantitatively.

For exact numerical evaluation, however, it is preferable not to rely on the small-$\delta$ expansion. The exact branch means are
\begin{equation}
\mathcal N_{\pm}=N_0\eta\exp\!\left[-\frac{(u\mp \delta/2)^2}{2\sigma^2}\right]
\left[1+V\cos\!\left(\kappa x_0+\phi_0\pm \frac{\kappa\delta}{2}\right)
\right].\label{eq:mupm}
\end{equation}
Since $\theta$ enters only through $u=x_0-\theta$, one obtains
\begin{equation}
\partial_\theta \mathcal N_{\pm}=\frac{u\mp \delta/2}{\sigma^2}\,\mathcal N_{\pm}.
\end{equation}
The corresponding exact Fisher information for the two independent Poisson branch counts is thus
\begin{align}
\FI_{\mathrm{TRY}}(\theta)&=\frac{(\partial_{\theta}\mathcal N_+)^2}{\mathcal N_+}+\frac{(\partial_{\theta}\mathcal N_-)^2}{\mathcal N_-}\nonumber\\
&=\frac{(u-\delta/2)^2}{\sigma^4}\mathcal N_++\frac{(u+\delta/2)^2}{\sigma^4}\mathcal N_-.
\label{eq:FIgauss}
\end{align}
Equation~(\ref{eq:FIgauss}) is the form used in the numerical optimization over the probe spacing $\delta$ in Sec.~\ref{sec:numerics}. It contains the full dependence on the finite probe separation and therefore avoids any ambiguity about the range of validity of the differential approximation. In contrast, Eqs.~(\ref{eq:Sgauss}) and (\ref{eq:dSgauss}) are most useful for physical interpretation, because they display explicitly how the envelope-gradient and interference-gradient terms combine in the local quadrature regime.

\section{Reference normalization and positive-only implementation}
\label{app:reference}

In a practical experiment, the total launched source power may drift slightly between the positive and negative branches of the differential code. If uncorrected, such multiplicative fluctuations can bias the measured subtraction and partially obscure the desired differential dependence on the unknown parameter. A convenient remedy is to monitor the exposure-dependent source level with a reference detector.

Let $R_{\pm}$ denote the corresponding reference counts recorded during the positive and negative branches. One may then form the normalized differential signal
\begin{equation}
\widetilde S_{\mathrm{diff}}=\frac{\mathcal N_+}{R_+}-\frac{\mathcal N_-}{R_-}.
\label{eq:Snorm}
\end{equation}
To first order, this normalization suppresses the effect of branch-to-branch source-power fluctuations while preserving the intended differential sensitivity to $\theta$. In this sense, it offers a more robust implementation of the positive-only two-branch protocol discussed in Sec.~\ref{sec:protocol}.

This normalization step is conceptually similar to the reference-based correction used in differential ghost imaging \cite{Ferri2010}, where common-mode intensity fluctuations are removed by comparing bucket signals to a simultaneously monitored reference channel. The role of the reference is different here, however, because the present forward model is deterministic and interferometric rather than correlation based. The normalization therefore should be viewed as an experimental stabilization procedure, not as part of the essential sensing mechanism itself.

In the idealized analysis of the main text, we assume sufficient source stability that the simpler unnormalized choice $S_{\rm diff}=\mathcal N_+-\mathcal N_-$ is adequate. Appendix~\ref{app:reference} is included to show how the same differential protocol can be implemented more robustly in realistic settings where branch-to-branch power fluctuations are non-negligible.


\begin{thebibliography}{99}

\bibitem{Young1804}
T.~Young, The Bakerian Lecture: Experiments and calculations relative to physical optics, Philos.\ Trans.\ R.\ Soc.\ Lond.\ \textbf{94}, 1--16 (1804).

\bibitem{BornWolf}
M.~Born and E.~Wolf, \textit{Principles of Optics}, 7th ed.\ (Cambridge University Press, Cambridge, 1999).

\bibitem{Goodman}
J.~W.~Goodman, \textit{Introduction to Fourier Optics}, 4th ed.\ (W. H. Freeman, New York, 2017).

\bibitem{MandelWolf}
L.~Mandel and E.~Wolf, \textit{Optical Coherence and Quantum Optics}
(Cambridge University Press, Cambridge, 1995).

\bibitem{Wen2025TRY}
J.~Wen, Time-reversed Young's experiment: Deterministic, diffractionless second-order interference effect, Opt.\ Commun.\ \textbf{597}, 132612 (2025).

\bibitem{Wen2026Hybrid}
J.~Wen, Metadata-conditioned coherence enables label-conditioned deterministic interference (submitted).

\bibitem{Shapiro2008}
J.~H.~Shapiro, Computational ghost imaging, Phys.\ Rev.\ A \textbf{78}, 061802(R) (2008).

\bibitem{Bromberg2009}
Y.~Bromberg, O.~Katz, and Y.~Silberberg, Ghost imaging with a single detector,
Phys.\ Rev.\ A \textbf{79}, 053840 (2009).

\bibitem{ErkmenShapiro2010}
B.~I.~Erkmen and J.~H.~Shapiro, Ghost imaging: from quantum to classical to computational, Adv.\ Opt.\ Photon.\ \textbf{2}, 405--450 (2010).

\bibitem{Ferri2010}
F.~Ferri, D.~Magatti, L.~A.~Lugiato, and A. Gatti,
Differential ghost imaging, Phys.\ Rev.\ Lett.\ \textbf{104}, 253603 (2010).

\bibitem{Bina2013}
M.~Bina, D.~Magatti, M.~Molteni, A.~Gatti, L.~A.~Lugiato, and F.~Ferri, 
Backscattering differential ghost imaging in turbid media, Phys.\  Rev.\ Lett.\ \textbf{110}, 083901 (2013).

\bibitem{Luo2012CPL}
K.-H.~Luo, B.-Q.~Huang, W.-M.~Zheng, and L.-A.~Wu, Nonlocal imaging by conditional averaging of random reference measurements, Chin.\ Phys.\ Lett.\ \textbf{29}, 074216 (2012).

\bibitem{Wen2012JOSAA}
J.~Wen, Forming positive-negative images using conditioned partial measurements from reference arm in ghost imaging, J.\ Opt.\ Soc.\ Am.\ A \textbf{29}, 1906--1911 (2012).

\bibitem{Morris2015}
P.~A.~Morris, R.~S.~Aspden, J.~E.~C.~Bell, R.~W.~Boyd, and M.~J.~Padgett, 
Imaging with a small number of photons, Nat.\ Commun.\ \textbf{6}, 5913 (2015).

\bibitem{PadgettBoyd2017}
M.~J.~Padgett and R.~W.~Boyd, An introduction to ghost imaging: quantum and classical, Philos.\ Trans.\ A\ Math.\ Phys. Eng. Sci. \textbf{375}, 20160233 (2017).

\bibitem{Edgar2019NP}
M.~P.~Edgar, G.~M.~Gibson, and M.~J.~Padgett, Principles and prospects for single-pixel imaging, Nat. Photonics \textbf{13}, 13--20 (2019).

\bibitem{Sun2020SPIReview}
M.-J.~Sun and J.~M.~Zhang, Single-pixel imaging and its applications in three-dimensional reconstruction: A brief review,
Sensors\ \textbf{19}, 732 (2019).

\bibitem{Gibson2020}
G.~M.~Gibson, S.~D.~Johnson, and M.~J.~Padgett, Single-pixel imaging 12 years on: a review, Opt.\ Express \textbf{28}, 28190--28208 (2020).

\bibitem{Helstrom}
C.~W.~Helstrom, \textit{Quantum Detection and Estimation Theory} (Academic Press, New York, 1976).

\bibitem{Kay}
S.~M.~Kay, \textit{Fundamentals of Statistical Signal Processing: Estimation Theory} (Prentice Hall, Upper Saddle River, 1993).

\bibitem{VanTrees}
H.~L.~Van Trees, \textit{Detection, Estimation, and Modulation Theory: Radar-Sonar Signal Processing and Gaussian Signals in Noise} (John Wiley \& Sons, Hoboken, 2001).

\bibitem{Tsang2016PRX}
M.~Tsang, R.~Nair, and X.-M.~Lu, Quantum theory of superresolution for two incoherent optical point sources, Phys.\ Rev.\ X \textbf{6}, 031033 (2016).

\bibitem{NairTsang2016}
R.~Nair and M.~Tsang, Far-field superresolution of thermal electromagnetic sources at the quantum limit,
Phys.\ Rev.\ Lett.\ \textbf{117}, 190801 (2016).

\bibitem{NairTsang2016OE}
R.~Nair and M.~Tsang, Interferometric superlocalization of two incoherent optical point sources, Opt.\ Express \textbf{24}, 3684--3701 (2016).

\bibitem{Yang2016}
F.~Yang, A.~Tashchilina, E.~S.~Moiseev, C.~Simon, and A.~I.~Lvovsky, Far-field linear optical superresolution via heterodyne detection in a higher-order local oscillator mode, Optica \textbf{3}, 1148--1152 (2016).

\bibitem{Tang2016}
Z.~S.~Tang, K.~Durak, and A.~Ling, Fault-tolerant and finite-error localization for point emitters within the diffraction limit, Opt. Express \textbf{24}, 22004--22012 (2016).

\bibitem{Tsang2018}
M.~Tsang, Subdiffraction incoherent optical imaging via spatial-mode demultiplexing, New J. Phys. \textbf{19}, 023054 (2017).

\bibitem{Backlund2018}
M.~P.~Backlund, Y.~Shechtman, and R.~L.~Walsworth,
Fundamental precision bounds for three-dimensional optical localization microscopy with Poisson statistics, Phys.\ Rev.\ Lett.\ \textbf{121}, 023904 (2018).

\bibitem{Napoli2019}
C.~Napoli, S. Piano, R.~Leach, G.~Adesso, and T.~Tufarelli, Towards supperresolution surface metrology: Quantum estimation of angular and axial separations, Phys.\ Rev.\ Lett.\ \textbf{122}, 140505 (2019). 

\bibitem{Francia1955}
G.~Toraldo di Francia, Resolving power and information, J.\ Opt.\ Soc.\ Am.\ \textbf{45}, 497--501 (1955).

\bibitem{Cox1986}
I.~J.~Cox and C.~J.~R.~Sheppard, Information capacity and resolution in an optical system, J.\ Opt.\ Soc.\ Am.\ A \textbf{3}, 1152--1158 (1986).

\bibitem{Gustafsson2000}
M.~G.~L.~Gustafsson, Surpassing the lateral resolution limit by a factor of two using structured illumination microscopy, J.\ Microsc.\ \textbf{198}, 82--87 (2001).

\bibitem{Gustafsson2009}
M.~G.~L.~Gustafsson, Nonlinear structured-illumination microscopy: Wide-field fluorescence imaging with theoretically unlimited resolution,
Proc. Natl. Acad. Sci. U.S.A. \textbf{102}, 13081--13086 (2005).

\bibitem{Saxena2015}
M.~Saxena, G.~Eluru, and S. S. Gorthi, Structured illumination microscopy, Adv.\ Opt.\ Photon. \textbf{7}, 241--275 (2015).

\bibitem{Schermelleh2019}
L.~Schermelleh, R.~Heintzmann, and H.~Leonhardt, A guide to super-resolution fluorescence microscopy,
J.\ Cell Biol.\ \textbf{190}, 165--175 (2010).

\bibitem{Hell1994}
S.~W.~Hell and J.~Wichmann, Breaking the diffraction resolution limit by stimulated emission: stimulated-emission-depletion fluorescence microscopy, Opt.\ Lett.\ \textbf{19}, 780--782 (1994).

\bibitem{Zhou2019}
Y.~Zhou, J.~Yang, J.~D.~Hassett, S.~M.~H. Rafsanjani, M.~Mirhosseini, A. N. Vamivakas, A. N. Jordon, Z. Shi, and R.~W.~Boyd, Quantum-limited estimation of the axial separation of two incoherent point sources, Optica \textbf{6}, 534--541 (2019).

\bibitem{Wadood2021}
S. A. Wadood, K. Liang, Y. Zhou, J. Yang, M. A. Alonso, X.-F. Qian, T. Malhotra, S. M. H. Rafsanjani, A. N. Jordon, R. W. Boyd, and A. N. Vamivakas, Experimental demonstration of superresolution of partially coherent light sources using parity sorting,
Opt. Express \textbf{29}, 22034--22043 (2021).


\end{thebibliography}
\end{document}